\documentstyle[twoside,fleqn,espcrc2,psfig]{article}

\newcommand{\AmS}{{\protect\the\textfont2
  A\kern-.1667em\lower.5ex\hbox{M}\kern-.125emS}}
\newcommand{\be}{\begin{equation}}
\newcommand{\ee}{\end{equation}}
\newcommand{\ben}{\begin{eqnarray}}
\newcommand{\een}{\end{eqnarray}}

\def\simgt{\rlap{\lower 3.5 pt\hbox{$\mathchar \sim$}}\raise 1pt \hbox {$>$}}
\def\simlt{\rlap{\lower 3.5 pt\hbox{$\mathchar \sim$}}\raise 1pt \hbox {$<$}}

\title{The Light Quark Masses with the Wilson Quark Action
using Chiral Ward Identities\thanks{presented by Y.~Kuramashi}}

\author{JLQCD Collaboration\\[2mm]
        S.~Aoki\address{Institute of Physics, University of Tsukuba,
        Tsukuba, Ibaraki 305, Japan},
        M.~Fukugita\address{Institute for Cosmic Ray Research, 
        University of Tokyo, Tanashi, Tokyo 188, Japan},
        S.~Hashimoto\address{Computing Research Center,
        High Energy Accelerator Research Organization (KEK),\\
        Tsukuba, Ibaraki 305, Japan},
        N.~Ishizuka$^{\rm a,}$\address{Center for Computational Physics, 
        University of Tsukuba, Tsukuba, Ibaraki 305, Japan},
        Y.~Iwasaki$^{\rm a,d}$,
        K.~Kanaya$^{\rm a,d}$
        Y.~Kuramashi\address{Institute of Particle and Nuclear Studies,
        High Energy Accelerator Research Organization (KEK),
        Tsukuba, Ibaraki 305, Japan},
        M.~Okawa$^{\rm e}$,
        A.~Ukawa$^{\rm a}$,
        T.~Yoshi\'e$^{\rm a,d}$
}
       
\begin{document}

\begin{abstract}

We present results for the light quark masses 
for the Wilson quark action obtained with the PCAC relation   
for the one-link extended axial vector current 
in quenched QCD at $\beta=5.9-6.5$.
This method leads to a remarkable improvement
of scaling behavior of the light quark masses
compared to the conventional method. 
We obtain ${\overline m}_l=3.87(37)$MeV
for the averaged up and down quark mass and 
${\overline m}_s=97(9)$MeV for the strange quark mass  
in the ${\overline{\rm MS}}$ scheme at $\mu=2$GeV.

\end{abstract}

\maketitle

\section{Introduction}

The chiral symmetry breaking term in the Wilson quark action 
causes large scaling violation effects of $O(a)$ 
for physical quantities in numerical simulations of lattice QCD. 
A manifestation of this effect is that the light quark masses 
for the Wilson action show a strong $a$ dependence 
at finite lattice spacings in contrast to a weak 
dependence for the case with the Kogut-Susskind (KS) action which 
retains U(1) chiral symmetry\cite{ukawa}.  
Uncertainties in a long extrapolation to the continuum limit needed 
for the Wilson case forms a part of the difficulty to settle the question  
whether the Wilson and KS quark actions give 
a consistent result in the continuum limit\cite{qmass_gb}. 
This leads us to reconsider 
the definition of quark mass for the Wilson action.

It is well known that the current quark mass   
defined by the PCAC relation for the Wilson action\cite{wi,ito} 
differs by $O(a)$ from the conventionally 
defined quark mass at finite lattice spacing.
We examine this point in more detail and find that 
the former definition applied with the one-link separated axial 
vector current that naturally arises in chiral Ward identities\cite{wi} 
leads to a significantly improved scaling behavior for 
the light quark masses.
  
Our calculations 
are carried out as a part of our simulation for the kaon 
$B$-parameter\cite{bk_w} with the Wilson quark action in quenched QCD.  
The plaquette gauge action is employed, and data are collected 
at $\beta=5.9-6.5$.  Point source quark propagators calculated 
for four values of the hopping parameter at each $\beta$ are 
used for this work.  Errors are estimated by the single elimination 
jackknife procedure for all measured quantities.

\section{Calculational method}

\begin{figure}[t]
\centering{
\hskip -0.0cm
\psfig{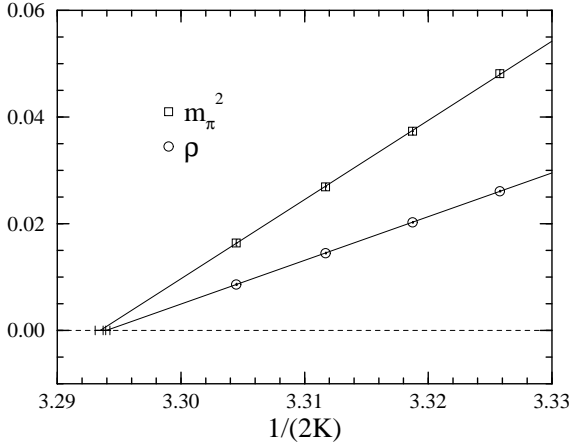}
\vskip -10mm  }
\caption{ $m_\pi^2$ and $\rho$ as a function of $1/(2K)$
at $\beta=6.3$.} 
\label{fig:rhop_63}
\vspace{-8mm}
\end{figure}

The conventional(CV) definition of quark mass 
is given by $m_q^{\rm CV} a=(1/K-1/K_c^{\rm CV})/2$ where
the critical hopping parameter 
$K_c^{\rm CV}$ is determined by extrapolating the pseudo scalar 
meson mass squared $m_{PS}^2$
linearly to $m_{PS}^2=0$ in the inverse hopping parameter 
$1/(2K)$. 
To extract the averaged up and down quark mass $m_l^{\rm CV}
=(m_u^{\rm CV}+m_d^{\rm CV})/2$ and the strange quark mass 
$m_s^{\rm CV}$   
we assume the following quark mass dependence for 
$m_{PS}$ and the vector meson mass $m_V$,
\ben
m_{PS}^2 &=& C_{PS}  (m_{q1}^{\rm CV} +m_{q2}^{\rm CV} )/2
\label{eq:mps_mq} \\
m_V  &=& m_V^0 +C_V(m_{q1}^{\rm CV} +m_{q2}^{\rm CV} )/2.
\een
The lattice spacing $a$ is fixed with $m_V^0$ using 
$m_\rho=770$MeV.
We employ the hadron mass ratio $m_\pi/m_\rho=0.1783$ 
to determine the averaged up and down quark mass 
$m_l^{\rm CV}$. 
The strange quark mass $m_s^{\rm CV}$ 
is estimated in two ways using 
$m_K/m_\rho=0.644$ and $m_\phi/m_\rho=1.323$. 
We convert $m_q^{\rm CV}$ into ${\overline m}_q^{\rm CV}$ 
defined in the $\overline{\rm MS}$
scheme at the scale $\mu=1/a$ 
using the tadpole-improved perturbative mass renormalization 
factor $Z_m(\mu a=1)$ evaluated with $\alpha_{\overline{\rm MS}}(1/a)$. 

The Ward identity (WI) method requires a calculation of
the $\rho$ parameter defined by the PCAC relation\cite{wi};
\be
2\rho(K)=\frac{\langle0|\nabla_\mu A_\mu^{{\rm ext},a}|\pi^a({\bf p}=0)\rangle}
{\langle0|P^{a}|\pi^a({\bf p}=0)\rangle},
\ee
We take the one-link extended axial vector current 
$A_\mu^{{\rm ext},a}$ since it is this current 
which naturally arises in the Ward identities.
We extract the $\rho$ parameter from the ratio
$\langle\nabla_4 A_4^{\rm ext}(t)P(1)\rangle/
\langle P(t)P(1)\rangle/4 
+\langle\nabla_4 A_4^{\rm ext}(T-t+2)P(1)\rangle/
\langle P(T-t+2)P(1)\rangle/4$, 
each two-point 
function projected to the zero spatial momentum, by fitting
a plateau as a function of $t$.
 
In Fig.~\ref{fig:rhop_63} a representative result for the 
$\rho$ parameter is plotted as a
function of $1/(2K)$ together with $m_\pi^2$.
We observe a clear linear behavior both for $\rho$ and $m_\pi^2$. 
The critical hopping parameter $K_c^{\rm WI}$ 
extracted from a linear extrapolation for $\rho$ 
is slightly different from $K_c^{\rm CV}$ for $m_\pi^2$
(see Table~\ref{tab:qmass} for numerical details).
We ascribe the discrepancy to uncertainties in 
the extrapolations of $\rho$ and $m_\pi^2$, because  
the critical hopping parameter obtained with the two definitions
should agree at each $\beta$.   
To avoid this problem, 
we define the bare quark mass for the WI method 
$m_q^{\rm WI}(1/a)$ by
\be
m_q^{\rm WI}(1/a)=\frac{\rho(K)}{(1/K-1/K_c^{\rm WI})/2}
m_q^{\rm CV}(1/a).
\ee
We convert quark masses defined on the lattice 
into those in the $\overline{\rm MS}$ scheme at the scale 
$\mu=1/a$ 
using the tadpole-improved perturbative renormalization factor 
$Z_{A^{\rm ext}}(\mu a=1)/Z_P(\mu a=1)$
evaluated with $\alpha_{\overline{\rm MS}}(1/a)$. 

We quote final results for ${\overline m}_q^{\rm CV,WI}$ at the 
scale $\mu=2$GeV which are obtained by a two-loop renormalization group 
running from $\mu=1/a$\cite{allton}.

\begin{figure}[t]
\centering{
\hskip -0.0cm
\psfig{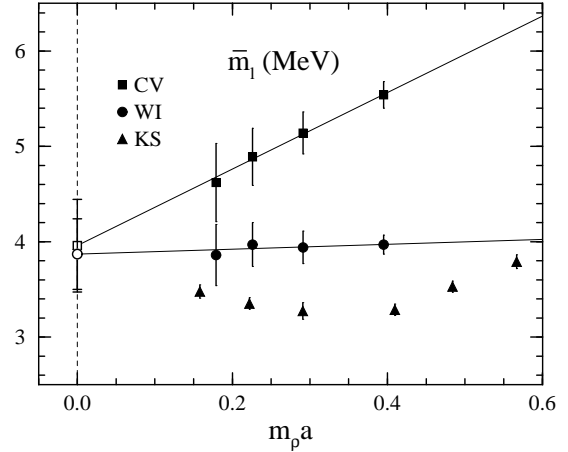}
\vskip -10mm  }
\caption{Averaged up and down quark mass 
in the ${\overline {\rm MS}}$ scheme 
at $\mu=2$GeV as a function of $a$.} 
\label{fig:qm_ud_cl}
\vspace{-8mm}
\end{figure}

\begin{table*}[t]
\vspace{-3mm}
\begin{center}
\caption{\label{tab:qmass} Critical hopping parameters 
and the light quark masses in the ${\overline {\rm MS}}$ scheme 
at $\mu=2$GeV.}
\begin{tabular*}{\textwidth}{@{}l@{\extracolsep{\fill}}llllll}\hline
        & input & $\beta$=5.9   & 6.1   & 6.3   & 6.5   & $a=0$ \\ 
\hline
$K_c^{\rm CV}$  & & $0.15986(3)$ & $0.15502(2)$     
                  & $0.15182(2)$ & $0.14946(3)$ \\ 
$K_c^{\rm WI}$  & & $0.15976(1)$ & $0.15497(1)$     
                  & $0.15179(1)$ & $0.14948(1)$ & \\ 
${\overline m}_l^{\rm CV}$ (MeV)     
                & $m_\pi/m_\rho$ & $5.54(14)$ & $5.14(22)$ 
                  & $4.89(30)$ & $4.62(41)$ & $3.96(49)$ \\
${\overline m}_l^{\rm WI}$ (MeV)     
                & $m_\pi/m_\rho$ & $3.97(10)$ & $3.94(17)$ 
                  & $3.97(23)$ & $3.86(32)$ & $3.87(37)$ \\
${\overline m}_s^{\rm CV}$ (MeV)     
                & $m_K/m_\rho$ & $139(3)$ & $129(5)$ 
                  & $123(8)$ & $116(10)$ & $100(12)$ \\
${\overline m}_s^{\rm WI}$ (MeV)     
                & $m_K/m_\rho$ & $100(3)$ & $99(4)$ 
                  & $100(6)$ & $97(8)$ & $97(9)$ \\
${\overline m}_s^{\rm CV}$ (MeV)     
                & $m_\phi/m_\rho$ & $184(13)$ & $149(17)$ 
                  & $151(24)$ & $120(23)$ & $$ \\
${\overline m}_s^{\rm WI}$ (MeV)     
                & $m_\phi/m_\rho$ & $132(9)$ & $115(13)$ 
                  & $122(20)$ & $99(19)$ & $$ \\
\hline
\end{tabular*} 
\end{center}
\vspace{-5mm}
\end{table*}

\section{Results for light quark masses}

In Table~\ref{tab:qmass}
we summarize the results for the averaged up 
and down quark mass ${\overline m}_l$. The lattice spacing dependence of 
${\overline m}_l$ is shown in
Fig.~\ref{fig:qm_ud_cl}. 
The results with the WI method show a remarkably flat behavior.
This allows a reliable extrapolation to the continuum limit 
by a linear function in $a$, with which 
we find ${\overline m}_l^{\rm WI}=3.87(37)$MeV at $a=0$. 
In contrast a large scaling violation effect is seen 
for the results with the CV method. 
While a linear extrapolation yields a consistent result 
${\overline m}_q^{\rm CV}=3.96(49)$MeV in the continuum limit, 
systematic uncertainties due to a long extrapolation are quite large.
Finding a theoretical explanation as to why finite $a$ corrections 
are so small for quark masses with the WI method would be 
an interesting problem\cite{kura}.

In Fig.~\ref{fig:qm_ud_cl} we also plot the KS
results\cite{yoshie} for comparison.
They show small scaling violations and
are systematically smaller than the results for the WI method.
However, we cannot conclude at this stage whether the Wilson results are
inconsistent with the KS results. 
The one-loop correction in the mass renormalization factor 
for the KS quark action is $50\%-100\%$ depending on the lattice spacing 
for the results in Fig.~\ref{fig:qm_ud_cl}, 
which suggests that higher order corrections might be large. 

Our results for the strange 
quark mass ${\overline m}_s$ 
determined from $m_K$ and $m_\phi$ are listed in Table~\ref{tab:qmass}.
We find that results with the WI method are quite flat also in this case.

Estimates of ${\overline m}_s$ using $m_K$ are expected
to satisfy ${\overline m}_s\approx 25{\overline m}_l$ for each $\beta$
if we assume the functional form (\ref{eq:mps_mq}), since then 
${m_K^2}/{m_\pi^2}=({{\overline m}_l+{\overline m}_s})/
({2{\overline m}_l})\approx 13$.  
Making linear extrapolations in $a$ for the results 
of the WI and the CV method we find a mutually consistent result in the 
continuum limit:  
${\overline m}_s^{\rm WI}=97(9)$MeV and 
${\overline m}_s^{\rm CV}=100(12)$MeV.

For the alternative determination of ${\overline m}_s$  
with the use of $m_\phi$,   
the discrepancy between the WI and the CV results reduces
toward the continuum limit.
However, large errors originating from those of the vector meson mass 
hinder us from reliably extrapolating 
the results to the continuum limit.

This work is supported by the Supercomputer
Project (No.~97-15) of High Energy Accelerator Research Organization (KEK),
and also in part by the Grants-in-Aid of
the Ministry of Education (Nos. 08640349, 08640350, 08640404,
09246206, 09304029, 09740226).


\vspace*{-2mm}

\begin{thebibliography}{99}

\bibitem{ukawa} A.~Ukawa, Nucl. Phys. {\bf B}
(Proc. Suppl.) {\bf 30} (1993) 3.

\bibitem{qmass_gb} R.~Gupta and T.~Bhattacharya, 
Phys. Rev. {\bf D55} (1997) 7203.

\bibitem{wi} M.~Bochicchio {\it et al.}, Nucl. Phys. {\bf B262} (1985) 331.

\bibitem{ito} S.\ Itoh {\it et al.}, Nucl. Phys. {\bf B274} (1986) 33.

\bibitem{bk_w} JLQCD Collaboration, S.~Aoki {\it et al.}, 
hep-lat/9705035. 

\bibitem{allton} C.~R.~Allton {\it et al.}, 
Nucl. Phys. {\bf B431} (1994) 667. 

\bibitem{kura} Y.~Kuramashi, in progress.

\bibitem{yoshie} JLQCD Collaboration, S.~Aoki {\it et al.}, Nucl. Phys. {\bf B}
(Proc. Suppl.) {\bf 53} (1997) 209.
 
\end{thebibliography}
\end{document}